\documentclass[fleqn,10pt]{wlscirep}

\usepackage[english]{babel}
\usepackage{graphicx}
\usepackage{dcolumn}
\usepackage{bm}
\usepackage{color}
\usepackage{todonotes}
\usepackage{calrsfs}
\usepackage{physics}
\usepackage{siunitx}
\DeclareMathAlphabet{\pazocal}{OMS}{zplm}{m}{n}

\renewcommand\vec{\mathbf}
\newcommand{\onlinecite}[1]{\hspace{-1 ex} \nocite{#1}\citenum{#1}} 

\title{Excitonic structure and pumping power dependent emission blue-shift of type-II quantum dots}

\author[1,2,*]{Petr Klenovsk\'y}
\author[1,2]{Petr Steindl}
\author[2]{Dominique Geffroy}
\affil[1]{Central European Institute of Technology, Masaryk University, Kamenice 753/5, 62500~Brno, Czech~Republic}
\affil[2]{Department of Condensed Matter Physics, Faculty of Science, Masaryk University, Kotl\'a\v{r}sk\'a~2, 61137~Brno, Czech~Republic}

\affil[*]{klenovsky@physics.muni.cz}

\date{\today}

\begin{abstract}
In this work we study theoretically and experimentally the multi-particle structure of the so-called type-II quantum dots with spatially separated electrons and holes. Our calculations based on customarily developed full configuration interaction approach reveal that exciton complexes containing holes interacting with two or more electrons exhibit fairly large antibinding energies. This effect is found to be the hallmark of the type-II confinement.
In addition, an approximate self-consistent solution of the multi-exciton problem allows us to explain two pronounced phenomena: the blue-shift of the emission with pumping and the large inhomogenous spectral broadening, both of those eluding explanation so far.
The results are confirmed by detailed intensity and polarization resolved photoluminescence measurements on a number of type-II samples. 
\end{abstract}

\begin{document}

\flushbottom
\maketitle
\thispagestyle{empty}
\section*{Introduction}
Semiconductor quantum dots (QDs) are currently one of the most promising candidate systems for the realization of quantum cryptography protocols~\cite{Muller2014,Strauf2007} or quantum gates~\cite{Stevenson2006,Rodt} in the information technologies. However, so-far most of the research has been focused on so-called type-I QDs where both electrons and holes are confined inside the dot body. Notably, much theoretical and experimental work
has been aimed at zeroing the fine-structure splitting (FSS) of the excitonic doublet by various methods~\cite{Gerardot2007,kleDresden,Trotta}.
On the other hand, research on type-II QDs~\cite{Akahane2004,BRUNNER1994,Kim2003,Stracke2014} 
where electron and hole states are spatially separated has not experienced that much interest. This is particularly due to the smaller overlap of the quasiparticle wavefunctions leading to considerably reduced emission intensity and rate~\cite{Klenovsky10,KleJOPCS,Hsu,Nishikawa2012}. Moreover, the emission from type-II QDs is considerably and inhomogeneously broadened~\cite{LiuSteer,Liu} compared to type I, and a large blue-shift of the emission energy with increasing laser pumping has been observed~\cite{Jin,UlloaHomogSRL}. 
Apart from a rich physics, the spatial separation of the carriers in systems with type-II confinement provides several advantages compared to type I. In particular, a naturally small FSS was predicted to occur, see Ref.~\onlinecite{Krapek2015},
as well as molecular-like states~\cite{Klenovsky10,KleJOPCS,KrapekNottingham} that might be used for the design of quantum gates, see Ref.~\onlinecite{Klenovsky2016}. As a result, type-II QDs can be used in quantum information technology without the need for elaborate post-processing of the quantum states.

Although we have recently demonstrated a method to determine the vertical position of the hole wavefunction in real samples~\cite{Klenovsky2015}, more obstacles associated with type-II QDs remain. In particular, because of the inhomogenous broadening of the spectral bands there has been only a small amount of experimental~\cite{Matsuda2007} or theoretical~\cite{RORISON1993,Miloszewski2014} studies of the excitonic structure of type-II QDs. Furthermore, there is still an ongoing discussion about the physical nature of the emission energy blue-shift with pumping~\cite{Gradkowski2012,Young2014,Hatami1998,Jo2012}. 

In this paper we address the aforementioned issues.
We provide the theoretical description of the (multi-)excitonic structure of type-II QDs using a Full Configuration Interaction (CI) method developed for this purpose, as well as the experimental observation of the recombination of multi-particle states by intensity and polarization resolved photoluminescence (PL) measurements. Finally, we present a model, based on approximate self-consistent CI, of the pumping dependent energy blue-shift of the individual excitonic transitions seen in our PL experiments, and provide its physical interpretation. 

Although our results are general, we have chosen GaAsSb capped InAs QDs in GaAs matrix as the test system. It is favored because this system allows for continuous change of the type of confinement by varying the Sb content in the GaAsSb capping layer (CL)~\cite{Klenovsky10}.
%
%
%
%
\section*{Theory}
We have studied excited states involving several interacting quasi-particles using 
the full configuration interaction technique~\cite{helgaker2008molecular,Schliwa2009,Stier2001,Zielinski2010}. In this framework, the stationary Schr\"{o}dinger equation to be solved reads
\begin{equation}
\label{eq:CISchroedinger}
\hat{H}^M\left|M\right>=M\left|M\right>,\,\,M\equiv X^0, X^+, X^-, XX^0\dots,
\end{equation}
where $M$ is the eigenenergy of the (multi-)excitonic state $\left|M\right>$ corresponding to $N_a$ and $N_b$ i.e.~the numbers of particles $a$ and $b$, respectively, where $a,b\in\{e,h\}$ with $e$ standing for electron and $h$ for hole. The complexes investigated in this work are the neutral exciton $X^0$ ($N_e=1$, $N_h=1$), the positive trion $X^+$ ($N_e=1$, $N_h=2$), the negative trion $X^-$ ($N_e=2$, $N_h=1$), and the neutral biexciton $XX^0$ ($N_e=2$, $N_h=2$). We look for solutions of Eq.~(\ref{eq:CISchroedinger}) in the form
\begin{equation}
\label{eq:CIWavefunction}
\left|M\right>=\sum_{m}\eta_{m}\left|D^M_{m}\right>
\end{equation}
where $m$ runs over all configurations corresponding to the given $M$. $\left|D^M_m\right>$ is the corresponding Slater determinant, which involves the single-particle states $\{ \psi_i\}, i \in S_n$ (so that the cardinal of $S_n$ verifies $\mathbf{card}(S_n) = N_a + N_b$). We solve variationally for the coefficients $\eta_m$, i.e. we solve the system of equations $\sum_m\left<D^M_n\right|\hat{H}^M\left|D^M_m\right>\eta_m=M\eta_n$, under the normalization constraint $\sum_n |\eta_{n}|^2=1$.

The matrix elements of the Hamiltonian $\hat{H}^M$ in the basis of the Slater determinants are given by (see the Supplementary Section SI online for details of the derivation, and for the expression of the Coulomb matrix element $V_{ij,kl}$)
\begin{equation}
  \begin{split}
\label{eq:CIHamiltonian}
\hat{H}^M_{nm} &\equiv \matrixel{D^M_n}{\hat{H}^M}{D^M_m}  \\
&=   \begin{cases}
    \mathcal{E}_n^{(e)}-\mathcal{E}_n^{(h)} 
    + \dfrac{1}{2}\sum\limits_{i,j\in S_n} &(V_{ij,ij} - V_{ij,ji})
     \text{  if $n = m$}\\
      \dfrac{1}{2} \sum\limits_{j\in S_n} V_{ij,kj} - V_{ij,jk} & \text{if $D^M_n$ and $D^M_m$ differ by one single-particle state: $\ket{D^M_n} \propto c^\dagger_i c_k \ket{D^M_m}$ } \\
      \dfrac{1}{2} (V_{ij,kl} - V_{ij,lk}) & \text{if $D^M_m$ and $D^M_n$ differ by two single-particle states: $\ket{D^M_n} \propto c^\dagger_i c^\dagger_j  c_k c_l \ket{D^M_m}$ , $k<l$},
    \end{cases}
  \end{split}
\end{equation}
where $\mathcal{E}^{(e)}_n$ ($\mathcal{E}^{(h)}_n$) is the sum of the energies of the occupied single-particle electron (hole) states in the determinant $D^M_n$.

The oscillator strength of the interband optical transition $F^M_{fi}$ between the $i$-th and $f$-th eigenstate of the excitonic complex $M$, respectively, is found using Fermi's golden rule~\cite{Zielinski2010}
\begin{equation}
\label{eq:CIOscStrength}
F^M_{fi}=\left|\left<M_f\left|\hat{P}\right|M_i\right>\right|^2,
\end{equation}
where $\left|M_f\right>$ is the final state after recombination of one electron-hole pair in $\left|M_i\right>$. The operator $\hat{P}$ is defined by
\begin{equation}
\label{eq:CIOscStrengthSP}
\hat{P}=\sum_{rp}\left<\psi_{e_r}\left|\vec{e}\cdot\hat{\vec{p}}\right|\psi_{h_p}\right>.
\end{equation}
Here $\psi_e$ and $\psi_h$ are the single-particle wavefunctions for electrons and holes, respectively, $\vec{e}$ is the polarization vector, and $\hat{\vec{p}}$ the momentum operator. For more information of the implementation of CI used in our calculations, see the Supplementary Section SI online.
%
%
%
%
%
\section*{Results}

\subsection*{Excitonic structure of type-II QDs}

\begin{figure}[!ht]
\begin{center}
\includegraphics[keepaspectratio=true,width=0.9\textwidth]{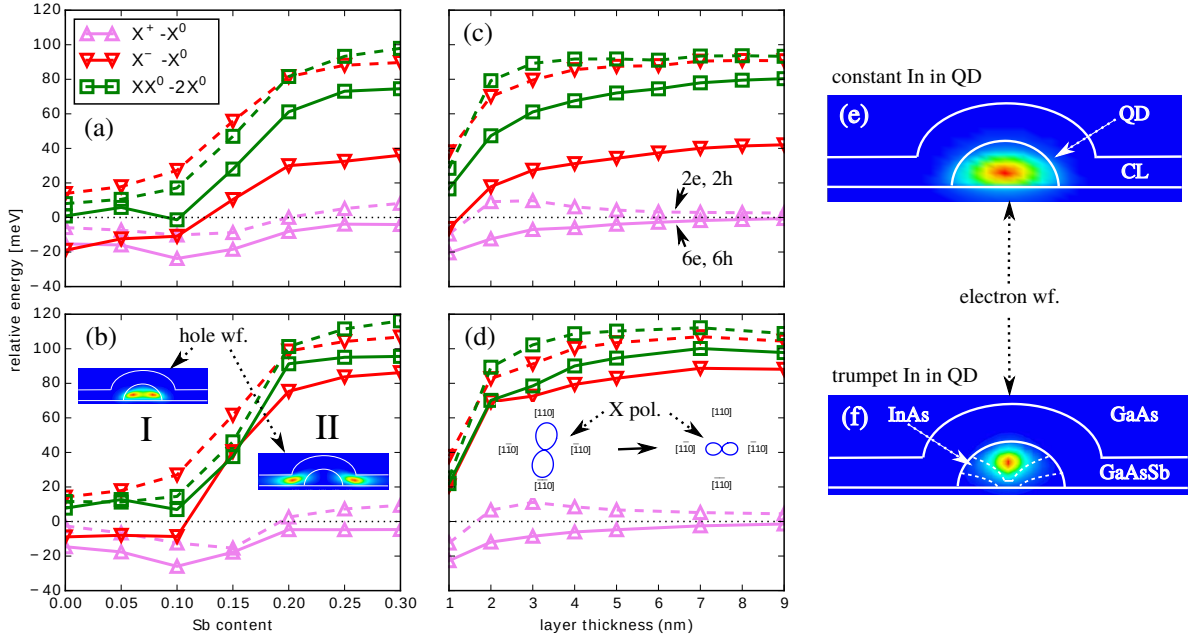}
\end{center}
\caption{(color online) (a) [(b)] Energies of the complexes $X^+$, $X^-$, and $XX^0$ with respect to that of $X^0$ as functions of Sb content for constant [trumpet] In composition in QD, and $d=$5~nm. Panel (c) [(d)] shows the same quantities as functions of $d$ for a fixed Sb content of 0.24. The points connected with broken curves correspond to calculations with 2 electron and 2 hole single-particle basis states while those connected by the full curves represent the results obtained for 6 by 6 basis. The labels I and II in (b) represent the type of confinement and the insets the corresponding $(1\bar{1}0)$ plane cuts of the hole probability densities. The inset of (d) shows the polarization of the emission of $X$ for thin and thick capping layers. Panels (e) and (f) show the simulated structures and the probability density of the electron wavefunction in QD for constant and trumpet In distribution, respectively.
\label{fig:calc_CIDep}}
\end{figure}

We have calculated the (multi-)excitonic structure of InAs/GaAsSb/GaAs QDs by the CI method, outlined in the previous section, coded in the Python 2.7 programming language~\cite{Python}. 
The single-particle basis states were obtained from the Nextnano++ simulation package~\cite{next} that employs the envelope function approximation based on the eight-band $\vec{k}\cdot\vec{p}$ method. The simulated structure was a lens-shaped InAs QD with height of 4~nm and basis diameter of 16~nm. The In distribution in the QDs was either constant at 0.6, or trumpet-like~\cite{Offermans,Ulloa}. The thickness $d$ and the Sb composition of the CL were varied during the calculations. For a side view of the structure, see panels (e) and (f) of Fig.~\ref{fig:calc_CIDep}. The physical properties of the simulated QDs were chosen so that the energy of $X^0$ approximately matches our experimental data, discussed in the next section.


In Figure~\ref{fig:calc_CIDep} (a) and (b) we show the calculated Sb content dependencies of the energy difference of $X^+$, $X^-$, and $XX^0$ from $X^0$.
It can be clearly seen that the transition from type-I to type-II confinement is associated with $XX^0$ and $X^-$ becoming significantly anti-binding while $X^+$ remains binding for all Sb contents. Depending on the shape and on the In composition profile of the type-II QD, $XX^0-2X^0$ varies from 80~meV for a lens-shaped QD with spatially constant In content~\cite{KleJOPCS,Klenovsky2015}, up to 200~meV for 
a pyramidal QD with trumpet In profile~\cite{Klenovsky10}.

These trends can be understood based on the fact that for type-II structures holes are located in CL whereas electrons in QD body, see the inset of Fig.~\ref{fig:calc_CIDep}~(b) for the hole, and panels (e) and (f) for the electron wavefunctions~\cite{Klenovsky10}. Because of that, the electron-hole attractive Coulomb interaction $J_{eh}$ is significantly reduced while the electron-electron $J_{ee}$ and hole-hole $J_{hh}$ repulsive potentials are either not changed at all or reduced much less. Since $XX^0-2X^0\approx J_{ee}+J_{hh}-2J_{eh}$ and $X^--X^0\approx J_{ee}-J_{eh}$, and considering that the wavefunctions of electrons do not change much with the Sb content~\cite{Klenovsky10}, one concludes that the dependencies are mainly governed by the reduction of $J_{eh}$, and one indeed expects that 
those should share a similar energy increase. On the other hand, changes in $X^+-X^0\approx J_{hh}-J_{eh}$ should be much smaller since the hole wavefunction tends to be segmented for type-II confinement and becomes more spatially spread in CL. This implies that $J_{hh}$ is reduced alongside with $J_{eh}$ for larger Sb. 
 
The binding nature of $X^+$ is, however, the result of the dynamic correlation characterized by a potential $E_{\mathrm{corr}}$. While the effect of $E_{\mathrm{corr}}$ on the studied complexes for QDs with trumpet In composition is generally not very significant [see Fig.~\ref{fig:calc_CIDep}~(b) and (d)], it can become as large as $-50$~meV for QDs with constant In content [see Fig.~\ref{fig:calc_CIDep}~(a)]. In all cases, $E_{\mathrm{corr}}$ tends to scale as $E_{\mathrm{corr}}(X^+)<E_{\mathrm{corr}}(XX^0)<E_{\mathrm{corr}}(X^-)$ in type-II QDs. 
The difference in the impact of $E_{\mathrm{corr}}$ on the $X^-$ and $X^+$ complexes is consistent with the different spatial spread of the single-particle electron and hole wavefunctions, respectively, which is smaller for the former. Similarly, QDs with a constant In composition behave as wider effective quantum wells for electrons, compared with their trumpet-like counterparts. This allows for an energetically more favorable spatial distribution of the single-particle wavefunctions and, thus, a larger reduction of $J_{ee}$ by $E_{\mathrm{corr}}$. Schliwa and coworkers observed similar trends for InGaAs/GaAs type-I QDs~\cite{Schliwa2009}. 

The results of $XX^0-2X^0$, $X^--X^0$, and $X^+-X^0$ as functions of $d$ for a fixed Sb content of $0.24$ can be found in panels (c) and (d) of Fig.~\ref{fig:calc_CIDep}; the content was chosen so that the system generally corresponds to type-II confinement. However, for thin layers the energy of holes is too large for localization of the hole ground state in the CL and the system is type I~\cite{Klenovsky10}, accompanied by a small value of $XX^0-2X^0$. However, further increase of $d$ brings the system to type-II confinement, marked again by enormously anti-binding $XX^0$ and $X^-$. It can be concluded that a strongly anti-binding biexciton is a hallmark of the type-II systems.
\begin{figure}[!ht]
\begin{center}
\includegraphics[keepaspectratio=true,width=0.8\textwidth]{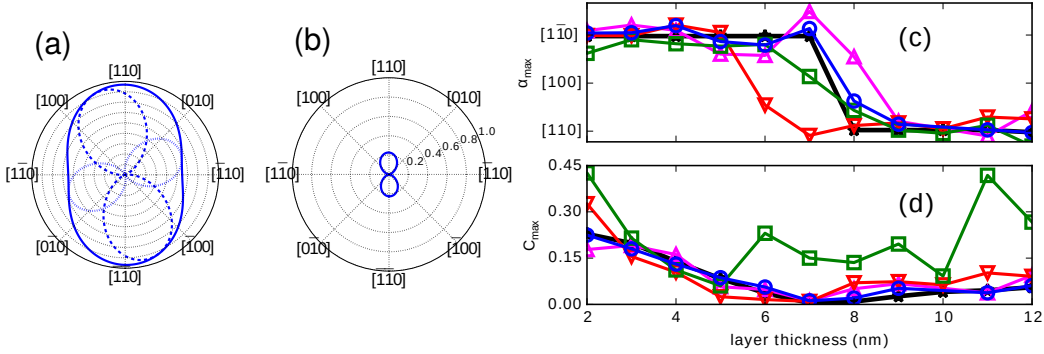}
\end{center}
\caption{(color online) (a) Polar graph of $F^X$ for two bright states lowest in energy (broken and dotted curve) and that for $\sum_{k=1}^2F^X_k$ (full curve). Panel (b) shows the polar graph of the degree of polarization $C(\alpha)$ for $\sum_{i=1}^4F^X_i$ (see text). Polarization azimuth $\alpha_{\mathrm{max}}$ and the maximum degree of polarization $C_{\mathrm{max}}$ as functions of CL thickness $d$ are given in panels (c) and (d), respectively. The data are plotted for $X$ (blue rings), $X^+$ (magenta upward triangles), $X^-$ (red downward triangles), and $XX$ (green squares). The multi-particle states were obtained using CI with 6 single-particle electron and 6 hole wavefunctions. For comparison we include the data taken from Ref.~\onlinecite{Klenovsky2015} (black stars).
\label{fig:calc_CIPol}}
\end{figure}

Moreover, the increase of $d$ induces a change in the vertical position of the hole wavefunction from the base of the QD to a location above its apex. This is connected with a change in the orientation of the emission polarization, and allows for the determination of the vertical position of the hole from polarization resolved PL measurements, see Ref.~\onlinecite{Klenovsky2015}. This effect is preserved also for states calculated by CI, see inset of Fig.~\ref{fig:calc_CIDep}~(d) and Fig.~\ref{fig:calc_CIPol}. However, one needs to calculate the emission of $X$ as $\sum_kF^X_k$ where $k$ runs over two bright excitonic eigenstates lowest in energy. The method is explained in panels (a) and (b) of Fig.~\ref{fig:calc_CIPol}. In (a) we show the polar graph of $F^X$ for the two lowest states of $X^0$, along different crystallographic directions. These are polarized linearly due to the structural anisotropy of the QDs and perpendicularly to each other, being oriented either close to $[110]$ or $[1\bar{1}0]$. The excitonic state with larger $F^X$ is oriented along the former direction and so is the polarization anisotropy of $\sum_{k=1}^2F^X_k$ for the exemplar QD. We characterize this, similarly to Ref.~\onlinecite{Klenovsky2015}, by a degree of polarization $C$ defined by 
\begin{equation}
\label{eq:degree_pol_definiton}
C(\alpha)=\frac{F(\alpha)-F_{\mathrm{min}}}{F_{\mathrm{max}}+F_{\mathrm{min}}},
\end{equation}
where $\alpha$ is the angle corresponding to the crystallographic direction in the plane of the sample; $F_{\mathrm{max}}$ and $F_{\mathrm{min}}$ are the maximum and minimum oscillator strengths, respectively. The maximum degree of polarization $C_{\mathrm{max}}$ occurs in direction given by an angle $\alpha_{\mathrm{max}}$.
We note that the reason for monitoring the polarization anisotropy of $\sum_{k=1}^2F^X_k$ instead of that of $F^X$ is due to the fact that we study ensemble PL in this work.

In panels (c) and (d) of Fig.~\ref{fig:calc_CIPol} we give $C_{\mathrm{max}}$ and $\alpha_{\mathrm{max}}$, respectively, for $X^0$, $X^+$, $X^-$, and $XX^0$ as functions of $d$.
It can be clearly seen that the polarization properties of the sum of multi-particle states mostly mimic that for $X^0$. This of course reflects the fact that on performing $\sum_kF^M_k$ the fine-structure of excitonic complexes is lost. We use this property to determine the origin of the multi-excitonic complexes in our PL measurements presented in the next section.

\subsection*{Experiment}

Four samples of InAs QDs with GaAsSb CL and a sample of InGaAs/GaAs QDs, were fabricated by the low pressure metal-organic vapor phase epitaxy (MOVPE)~\cite{Hospodkova2010surfSci}. Detailed information about the sample preparation might be found in Ref.~\onlinecite{Klenovsky2015}. We note here only that in order to obtain statistically significant variability of QDs differing in sizes, shapes, CL thicknesses and compositions, the growth was performed on a non-rotating susceptor.

The PL measurements were performed using a NT-MDT Ntegra-Spectra spectrometer. The samples were positioned in the cryostat, cooled to liquid nitrogen temperature (LN2) and pumped by a solid-state laser with a wavelength of~785~nm. The maximum laser power on the sample surface was 5~mW on the 150~$\mu$m$^2$ area, and was varied by a tunable neutral density (ND) filter.
The polarization of the PL was analyzed by a rotating achromatic half-wave plate followed by a fixed linear polarizer. The PL signal was dispersed by a 150 grooves/mm ruled grating and detected by the InGaAs line-CCD camera, cooled to minus~\SI{90}{\degreeCelsius}. In every experiment, PL light was collected from a large number of QDs ($\sim$3000).

\begin{figure}[!ht]
\begin{center}
\includegraphics[keepaspectratio=true,width=0.8\textwidth]{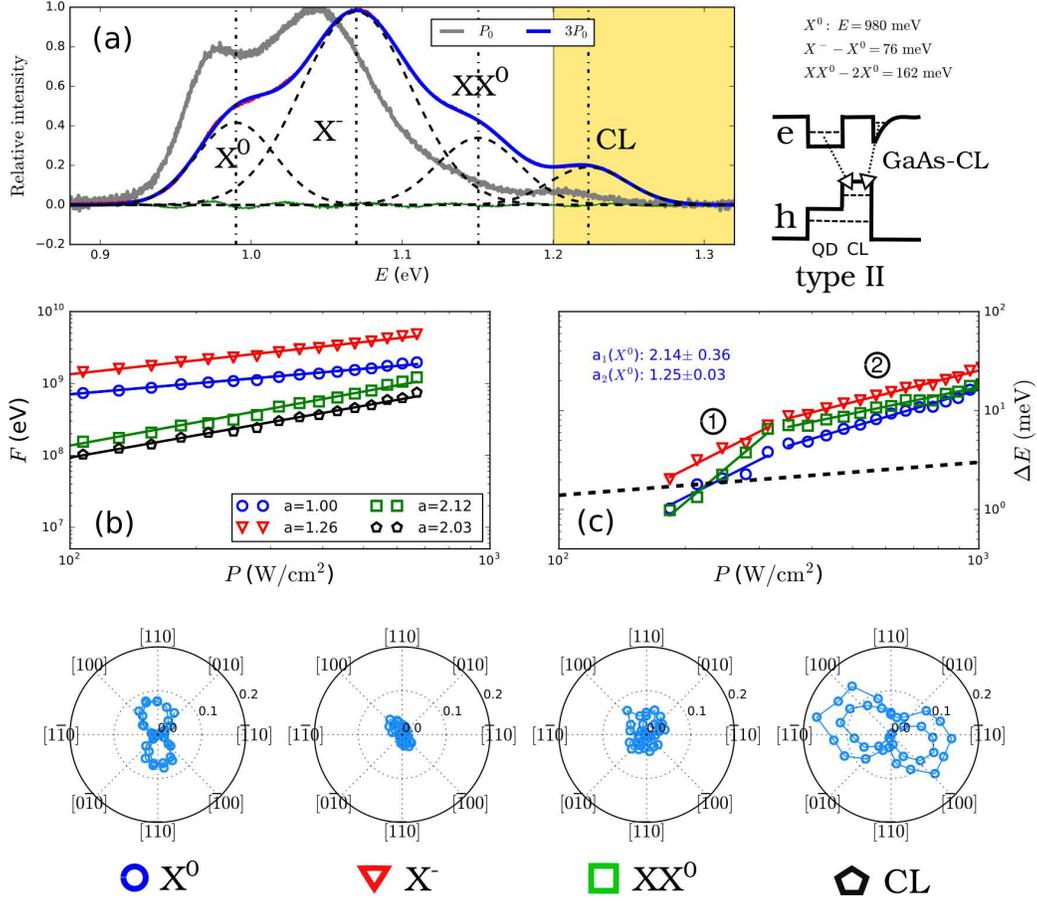}
\end{center}
\caption{(color online) (a) PL spectra of GaAsSb capped InAs type-II QDs measured for two pumping powers $P$ (grey and blue curves). The fit by the sum of Gaussian curves is shown for $3P_0$ (blue curve) and the individual bands corresponding to different multi-excitonic transitions are shown by broken curves. The difference between data and fit is represented by green curve. The dotted vertical lines indicate the energies of the bands for $3P_0$. The yellow shaded part of the graph corresponds to the recombination between bulk GaAs and CL. The inset next to panel (a) gives the spectral positions of of the three bands and a schematic band diagram of the recombination processes (not in scale). In (b) we show the $P$-dependence of the oscillator strength $F$ of the identified bands in log-log scale and their fits by linear lines, respectively, for $X^0$ (blue circles), $X^-$ (red triangles), $XX^0$ (green squares), and that for the transition between bulk GaAs and CL (black pentagons). The slopes $a$ (i.e. exponents in the linear plots) of the fitted lines are given in the inset of panel (b), for clarity they were normalized so that that $a=1$ for $X^0$. Panel (c) depicts the change of the emission energy $\Delta E$ with $P$ in log-log scale. Except of GaAs-CL transition which was omitted, the labels are the same as in (b) and the data were fitted by two linear functions in segments 1 and 2 (see text). The fitted slopes (i.e. exponents of the dependencies) $a_1$ and $a_2$ for $X^0$ are given in the inset. The $\Delta E\sim P^{1/3}$ dependence of Ref.~\onlinecite{Hatami1998} is shown by broke curve. The polar graphs at the bottom show $C(\alpha)$ of individual identified bands.
\label{fig:PL_typeII}}
\end{figure}
We show our experimental results for one sample with type-II QDs in Fig.~\ref{fig:PL_typeII}. We have measured PL spectra for a set of values of the laser pumping power $P$. For each of these values also the polarization anisotropy $C(\alpha)$ defined by Eq~(\ref{eq:degree_pol_definiton}) was obtained. The spectra were fitted by a sum of Gaussian functions and to each of those an appropriate recombination channel was assigned, see Fig.~\ref{fig:PL_typeII}~(a). The assignment was based on (i) the exponent $a$ of $F=b\times P^a$, see Fig.~\ref{fig:PL_typeII}~(b), (ii) the azimuth $\alpha_{\mathrm{max}}$, see the polar graphs at the bottom of Fig.~\ref{fig:PL_typeII}, and (iii) comparison with theoretical predictions.
The type of confinement was determined, as it is usually done~\cite{LiuSteer,Ulloa2012}, by observation of the blue-shift of the normalized PL spectra with $P$ (or its absence), see panel (a).

We have clearly identified the recombination of $X^0$, $X^-$, and $XX^0$ in our spectra. The measured energy separations of the latter two from $X^0$ are $X^--X^0=76$~meV and $XX^0-2X^0=162$~meV, i.e. considerably shifted to higher energies as predicted by our theory, see Fig.~\ref{fig:calc_CIDep}. As expected, the band $X^+$ cannot be distinguished from $X^0$ in our PL measurements since 
$X^+-X^0$ is much smaller than the inhomogenous spectral broadening of the emission bands from GaAsSb capped InAs QDs. 

The band with the largest energy $E$ [shaded part in Fig.~\ref{fig:PL_typeII}~(a)] is attributed to the recombination between electrons confined due to strain at the interface between bulk GaAs, and holes within the GaAsSb CL, see also inset of Fig.~\ref{fig:PL_typeII}. 
This transition is purely of type-II nature and exhibits a very large blue-shift with increasing $P$. A similar recombination pattern has been observed previously for InAsSb QDs~\cite{Mazur2012} and is also responsible for the no-phonon emission from SiGe/Si QDs~\cite{SiGeKlenovsky}.

In Fig.~\ref{fig:PL_typeII}~(c) we show in log-log scale the energy shift of the PL bands (excluding GaAs-CL transition) with increasing $P$, i.e.~$\Delta E(P)=E^M(P)-E^M(P_{\mathrm{min}})$ where $P_{\mathrm{min}}$ is the lowest value of $P$ used in our measurements. In addition to a considerable blue-shift of the bands, which is as large as 30~meV for our values of $P$, it can be seen that (i) $\Delta E$ is different for different $M$ and (ii) there is 
an edge in the power dependencies meaning that $\Delta E(P)$ 
does not follow the simple form of $\Delta E=b\times P^a$ for all values of $P$ but that for each of the two segments 1 and 2 a different exponent 
$a_1=2.14\pm0.36$ and $a_2=1.25\pm0.03$ seem more appropriate. That is significantly different from the $\Delta E\sim P^{1/3}$ dependency predicted in Ref.~\onlinecite{Hatami1998}, which is displayed as a broken curve for comparison.
Moreover, there are differences between the values of the exponents for different complexes. We postpone the explanation of the pumping dependent blue-shift to the next section.

The polar graphs at the bottom of Fig.~\ref{fig:PL_typeII} show that the radiation due to the recombination of 
$M$s from the dots is polarized along [110] crystallographic direction. The fact that all $M$s share the same $\alpha_{\mathrm{max}}$ confirms our prediction shown in Fig.~\ref{fig:calc_CIPol}. The polarization along the [110] in turn means that the holes are located in the CL close to the QD base~\cite{Klenovsky2015} for this particular sample. Moreover, the measurements of $\alpha_{\mathrm{max}}$ allow us to clearly distinguish the GaAs-CL band from the recombinations originating from QD states 
owing to their perpendicular polarization. A similar method of assignment was used in earlier studies, see e.g.~Ref.~\onlinecite{Alonso-Alvarez2011a}.

%
%
%
%
\begin{figure}[!ht]
\begin{center}

\includegraphics[keepaspectratio=true,width=0.8\textwidth]{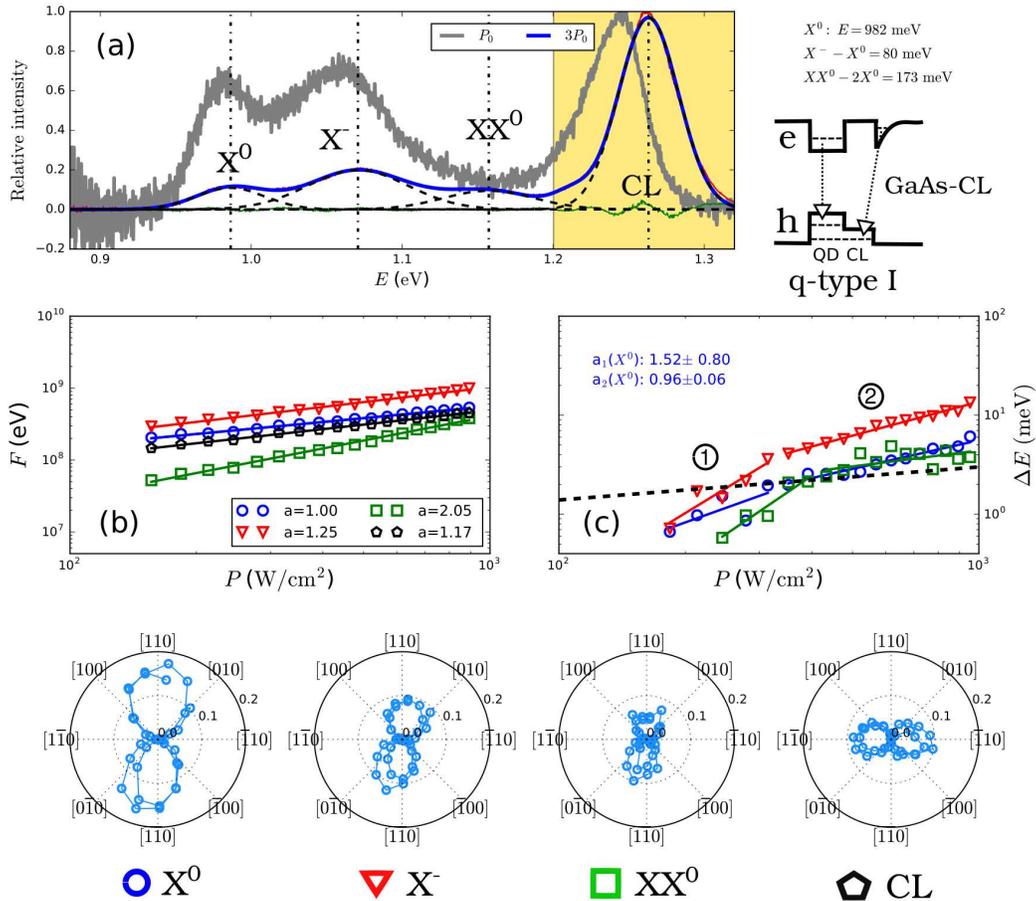}
\end{center}
\caption{(color online) PL spectra of GaAsSb capped InAs q-type-I QDs. The results are given in the same nomenclature as in Fig.~\ref{fig:PL_typeII}.
\label{fig:PL_qtypeI}}
\end{figure}
In Fig.~\ref{fig:PL_qtypeI} we show PL measurements of QDs with type-I confinement in very much the same way as in Fig.~\ref{fig:PL_typeII}. We again identify $X^0$, $X^-$, $XX^0$ and the GaAs-CL transition, respectively. Interestingly, the relative positions of $M$s are similar to type-II QDs, with $X^--X^0=80$~meV and $XX^0-2X^0=173$~meV. This is due to the fact that even a rather small increase in CL thickness results in a considerable increase of $X^--X^0$ and $XX^0-2X^0$, respectively, see Fig.~\ref{fig:calc_CIDep}. In fact, the type of confinement is not purely of type-I in GaAsSb capped InAs QDs for Sb contents different from zero, since even a slight increase of that lowers the confinement for holes in the CL with respect to GaAs~\cite{Klenovsky10}, see inset of Fig.~\ref{fig:PL_qtypeI}. Thus, the excited single-particle states tend to be partly localized in the CL, which results in a slight blue-shift of $X^0$ with increasing $P$ and also $a_1=1.52\pm0.80$ and $a_2=0.96\pm0.06$ are not equal to zero, respectively, see Fig.~\ref{fig:PL_qtypeI}~(c). We call this type of confinement ``quasi-type-I" (q-type-I). For results of the true type-I confinement represented by InAs/GaAs QDs for which no blue-shift of $X^0$ with $P$ is observed and $a\approx0$ see Supplementary Fig. S1 online. We note that slightly larger values of $X^--X^0$ and $XX^0-2X^0$ for type-II and q-type-I in Figs.~\ref{fig:PL_typeII} and~\ref{fig:PL_qtypeI}, respectively, compared to Fig.~\ref{fig:calc_CIDep} are probably due to the spatial inhomogeneity of the In distribution in the QDs, and of Sb in the CL, which were not considered in our theory. 

The polarization anisotropy of q-type-I QDs is again along [110], see bottom of Fig.~\ref{fig:PL_qtypeI}, and is larger than for type II in agreement with our theoretical predictions in Fig.~\ref{fig:calc_CIPol}~(d) for very thin CL. Again the GaAs-CL band is identified by its perpendicular polarization compared to the emission from QD bands. Note that the energy difference of that band from $X^0$ is larger compared with the type II. This is consistent with our interpretation, Cf. insets of Figs.~\ref{fig:PL_qtypeI} and~\ref{fig:PL_typeII}. Since the confinement for holes in the CL is larger for type I than for type II, the emission energy of GaAs-CL band is larger in type I. The energy of the GaAs-CL band also blue-shifts with pumping by a much larger amount than for the q-type-I QD bands and, thus, this structure represents a coexistence of type-I and type-II confinement~\cite{Ji2015}.

In order to make our results more general
we repeated the aforementioned measurements on 16 different QD samples. The results of those are given in the Supplementary Figs.~S2 for the exponent in $F=bP^a$, S3 for the mean emission energy shift, S4 for the mean $a$ in $\Delta E=bP^a$, and S5 for $\alpha_{\mathrm{max}}$, and $C_{\mathrm{max}}$, respectively.
%
%
%
%
%
%
%
%
\subsection*{Model of blue-shift}
\label{sec:model_blue_shift}

We comment here on the origin of the blue-shift of the emission energy $\Delta E$ with increasing $P$ in type II which we observe in our PL measurements. Currently, two competing hypotheses are put forward in this respect: the~``state-filling" model, stemming from the observation of the large radiative lifetime of the emission from type-II QDs~\cite{Liao2009,Nishikawa2012,Sato2012,Pavarelli2012,Young2014} which occurs due to the  smaller overlap between electron and hole wavefunctions~\cite{Klenovsky10,Hsu}. If the pumping rate exceeds the emission rate of the ground state transition, which is often the case in type-II QDs, a larger proportion of the radiative transitions between electronic levels higher in energy than the ground state results. This in total leads to the overall blue-shift of the emission~\cite{Gradkowski2012}. 
On the other hand, the so-called ``band-bending" hypothesis explains the blue-shift as a change of the confinement potential for holes close to the QD~\cite{LiuSteer,Jin,Hatami1998,Jo2012}. 
For the latter hypothesis the shift was characterized by a $\Delta E\sim P^{1/3}$ dependence in case of GaSb/GaAs QDs, see Refs.~\onlinecite{HATAMI1995,Hatami1998}, or as a change of the slope of one barrier in an infinite triangular quantum well~\cite{Kuokstis2002,Jo2012}. The fundamental difference between these two hypotheses lies in whether or not the energies of the quantized levels change with $P$ in type-II QDs. We now attempt to simulate and interpret the experimentally observed blue-shift.

\begin{figure}[!ht]
\begin{center}
\includegraphics[keepaspectratio=true,width=0.45\textwidth]{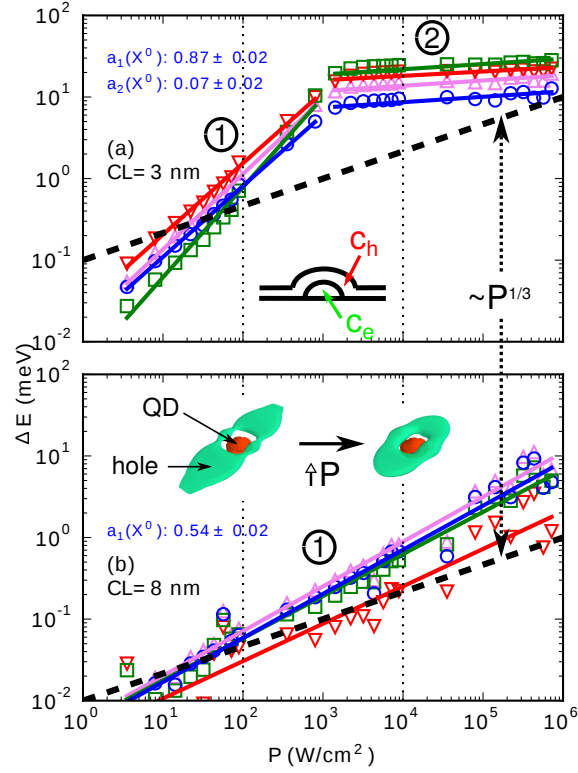}
\end{center}
\caption{(color online) Energy shift $\Delta E$ versus $P$ for CL thicknesses $d$ of (a) 3~nm and (b) 8~nm. The results are shown for $X^0$ (blue circles), $X^+$ (magenta upward triangles), $X^-$ (red downward triangles), and $XX^0$ (green squares). The inset in (a) depicts the background electron ($c_e$) concentration in QD and hole ($c_h$) in CL. The inset in (b) gives the the ground state hole probability densities for $d$=8~nm and two values of $P$. The wavefunctions are given as green isosurfaces 
encircling 50\% of the probability; the dot is marked as a red lens in the middle of the hole wavefunction. The parts of the wavefunction with largest probability density are oriented along the [110] direction. The numbers 1 and 2 mark different segments of the dependency similarly to Figs.~\ref{fig:PL_typeII} and \ref{fig:PL_qtypeI}, respectively. The fitted values of the exponents $a_1$ and $a_2$ for $X^0$ are given in the insets of both panels.
\label{fig:blueShift_model}}
\end{figure}
We first note that the model $\Delta E\sim P^{1/3}$
does not fit our experimental results presented in Figs.~\ref{fig:PL_typeII}~and~\ref{fig:PL_qtypeI}. 
By contrast, we follow an approach similar to that of Gradkowski~et~al., see Ref.~\onlinecite{Gradkowski2012}.
Since the electronic states in QDs are multi-particle in nature we proceed by employing a semi-self-consistent CI (SSCCI) approach to characterize the energy shift with $P$. In SSCCI we first define a certain concentration of background electron $c_e$ and hole $c_h$ charges, the former only in the QD body and the latter just in CL, respectively, reflecting the spatially indirect nature of charges in type-II system. The concentrations are defined in such a way that $c_h-c_e=0$, corresponding to a steady-state pumping. In other parts of the simulated structure (GaAs matrix) we set $c_h=c_e=0$, reflecting the fact that the recombination there is spatially direct. The presence of $c_e$ and $c_h$ results in a background electric potential in QD and CL, respectively.
We solve self-consistently the single-particle Schr\"odinger and Poisson equations for the system in the presence of these potentials.
The wavefunctions obtained in that way are then fed into our CI solver. We, however, set all diagonal matrix elements of $J_{eh}$ resulting from the CI calculation
to zero since we assume those to be already included in the single-particle energies obtained from the preceding self-consistent cycle.

On the other hand, the correspondence with the experiment is obtained by~\cite{Kuokstis2002}
\begin{equation}
\label{eq:conc_to_P_recalc}
P=\frac{10^6\gamma E_{l}}{\alpha}c_q^2,
\end{equation}
where $c_q=c_e=c_h$ is the number of electron-hole pairs generated by a laser light with the energy $E_l$ (in our case 1.58~eV), $\alpha$ is the absorption coefficient of the sample material ($2.0\times 10^4$~cm$^{-1}$ for GaAs~\cite{landoltbornstein}), and $\gamma$ is the radiative recombination coefficient ($7.0\times 10^{-10}$~cm$^3$s$^{-1}$ in the case of GaAs~\cite{landoltbornstein}).

The results of the blue-shift of multi-excitonic energies with $P$ calculated by SSCCI are presented in Fig.~\ref{fig:blueShift_model} for CL thicknesses of 3~and~8~nm. Clearly, our model correctly predicts the ``bending" of $\Delta E(P)$ in panels (c) of Figs.~\ref{fig:PL_typeII}~and~\ref{fig:PL_qtypeI}, and the presence of different slopes in the log-log graph [indicated by marks 1 and 2 in Fig.~\ref{fig:blueShift_model}~(a)]. The fitted exponents (slopes in log-log graphs) are $a_1=0.87\pm0.02$ and $a_2=0.07\pm0.02$ for $d=$3~nm, and $a_1=0.54\pm0.02$ for 8~nm, respectively. Such slopes are rather close to those in Fig.~\ref{fig:PL_typeII}. For type II the exponent in sector 1 decreases while that in sector 2 increases with increasing $d$, up to an approximate $\Delta E\sim P^{1/3}$ dependence reached for thick CLs. The blue-shift is accompanied by a change of the spatial distribution of the hole wavefunction~\cite{Gradkowski2012}, see insets of Fig.~\ref{fig:blueShift_model}~(b): holes are ``squeezed" towards the QD body and, thus, towards the electrons~\cite{Gradkowski2012}. 

The difference between the $a_1$ exponents for thinner and thicker CLs, respectively, can be qualitatively understood by inspecting the change in the lateral electron-hole dipole moment $p_{eh,xy}$ with $P$. We focus on the lateral plane only, where the anisotropy of the confinement potential for holes due to piezoelectricity $V^{\mathrm{piez}}_{xy}$ is strongest~\cite{Klenovsky10}. The total lateral potential for holes in type-II QDs might be written as $V^{\mathrm{total}}_{xy}=V^{\mathrm{bulk}}_{xy}-V^{\mathrm{piez}}_{xy}$, where $V^{\mathrm{bulk}}_{xy}$ is the energy of the holes in bulk semiconductor.
The spatial extent of the hole wavefunction is shifted towards that of the electrons, confined in QD body for type II, with increasing $P$. Thus, $V^{\mathrm{piez}}_{xy}$ is reduced, so that both $V^{\mathrm{total}}_{xy}$ and the single-particle transition energies increase. By noticing that $p_{eh,xy}$ is larger in the QDs with a thin CL than in those with a thick CL, we can infer that the rate of blue-shift of single-particle transition energies with $P$ should be larger for QDs with thinner CLs.

The mechanism just described can only be valid, however, up to some finite pumping power $P_{\mathrm{crit}}$: at some point, the drift of the hole towards the electrons is hampered by the confinement imposed by the QD body, lest the hole either collapses into QD
and the system transfers to the type-I confinement, or it is not bound in the structure anymore.  At this point, we should observe a reduction in the rate of the change of the dipole. Following this scenario, $P_{\mathrm{crit}}$ must be smaller for thinner CLs and should increase with $d$, see Supplementary Fig.~S6 for $d=5$~nm.
 
The meaning of $c_e$ and $c_h$ in our calculations is that of an average occupation of charge traps~\cite{Reimer2016} in QD and CL, respectively, changing from less than one to two electron-hole pairs per QD volume as $P$ is increased. In our understanding, there is probably an important contribution of trap-state filling effect to the emission blue-shift with pumping in type-II. This is consistent with the inhomogeneously broadened spectra of single type-II QDs observed elsewhere~\cite{Turck2000,Grydlik2015}, given the known link between trap filling and spectral broadening.
We note that non-broadened recombination from type-II QDs might be experimentally measured, e.g.~by using photon correlation Fourier spectroscopy~\cite{BrokmannOE2009,Bakulin2013}. Finally, our model of blue-shift of emission energy with $P$ is more consistent with the ``band-bending'' hypothesis presented above if the changes of the confinement potential of quasiparticles are due to filling of charge traps.

\section*{Conclusions}
We have studied the excitonic structure of type-II quantum dots both theoretically using the full configuration interaction method and experimentally by intensity and polarization resolved photoluminescence spectroscopy. We have found that the multi-particle complexes containing more electrons than holes in type II are typically considerably antibinding compared to type I, up to almost 200~meV for biexcitons. Furthermore, the polarization resolved photoluminescence measurements enabled us to discriminate spatially different recombination channels in our structures and revealed the coexistence of type-I and type-II confinement. Finally, the approximate self-consistent multi-particle calculations modeled our experimentally observed blue-shift of the emission with pumping in type II and provided an insight into the nature of that, being due to the filling of trap states, explaining also the reason for the large inhomogeneous spectral broadening of the emission bands of type-II QDs.
%
%
%
\bibliography{ref_pk_Utf8_paper}

\section*{Acknowledgments}
The authors thank Alice Hospodkov\'a for her help with the sample preparation and Dominik Munzar, Du\v{s}an Hemzal, and Vlastimil K\v{r}\'apek for insightful discussions. 
A part of the work was carried
out under the project CEITEC 2020 (LQ1601) with financial support
from the Ministry of Education, Youth and Sports of the Czech Republic
under the National Sustainability Programme II.

\section*{Author contributions}
P.K. developed the CI code and the theory, interpreted the experimental data and wrote the manuscript. P.S. performed the measurements and provided figures with experimental data. D.G. provided the formal description of the CI theory presented in the supplement as well as performed the language correction of the manuscript.

\section*{Additional Information}
{\bf Supplementary information} accompanies this paper at http://www.nature.com/srep\\

\noindent{\bf Competing financial interests:} The authors declare no competing financial interests.

\end{document}